\documentclass[aps,prl,twocolumn,showpacs,preprintnumbers,superscriptaddress]{revtex4}

\usepackage{graphicx}

\begin{document}


\def\coll {{\rm Coll.}}
\def\etal {{\it et~al.}}
\def\prep {{\it preprint}}
\def\EJP {{Eur.~J.~Phys..~}}
\def\NIM {{Nucl.~Instr.~Meth.~}}
\def\PRL {{Phys.~Rev.~Lett.~}}
\def\PR {{Phys.~Rev.~}}
\def\PL {{Phys.~Lett.~}}
\def\ZP {{Z.~Physik~}}

\def\thp	{$\Theta^+$}
\def\thc        {$\Theta^0_c$}
\def\ximm	{$\Xi^{--}$}
\def\xizz	{$\Xi^{0}_{3/2}$}
\def\xipp	{$\bar{\Xi}^{++}_{3/2}$}
\def\xim	{$\Xi^{-}$}
\def\xip	{$\bar{\Xi}^{+}$}
\def\xiq	{$\Xi_{3/2}$}
\def\xis	{$\Xi^{*}$}
\def\xiz	{$\Xi^0$}
\def\xisz	{$\Xi(1530)^0$}
\def\axisz	{$\bar{\Xi}(1530)^0$}
\def\xipic	{$\Xi^-\pi^-(\bar{\Xi}^+\pi^+)$}
\def\xipin	{$\Xi^-\pi^+(\bar{\Xi}^+\pi^-)$}

\def\pip	{$\pi^+$}
\def\pim	{$\pi^-$}
\def\ks		{K$^0_S$}
\def\kp         {K$^+$}
\def\km         {K$^-$}
\def\kz	{K$^0$}
\def\lam	{$\Lambda$}
\def\alam	{$\bar{\Lambda}$}
\def\sol        {{\it c}}
\def\solq       {{\it c$^2$}}
\def\MEV       {{MeV/\it c$^2$}}
\def\GEV       {{GeV/\it c$^2$}}
\newfont{\ord}{cmsy10 scaled 1000}    
\def\BR        {{\ord B}}
\def\BRD       {{\ord B}$\cdot$}
\def\BRS       {{\ord B}$\cdot$d$\sigma$/dy}
\def\UL        {UL(95\%)}

\preprint{DESY 04-148}

\title {
Limits for the central production of \thp\ and \ximm\ pentaquarks  in 920~GeV pA collisions 
} 

\affiliation{\it NIKHEF, 1009 DB Amsterdam, The Netherlands}
\affiliation{\it Department ECM, Faculty of Physics, University of Barcelona, E-08028 Barcelona, Spain}
\affiliation{\it Institute for High Energy Physics, Beijing 100039, P.R. China}
\affiliation{\it Institute of Engineering Physics, Tsinghua University, Beijing 100084, P.R. China}
\affiliation{\it Institut f\"ur Physik, Humboldt-Universit\"at zu Berlin, D-12489 Berlin, Germany}
\affiliation{\it Dipartimento di Fisica dell' Universit\`{a} di Bologna and INFN Sezione di Bologna, I-40126 Bologna, Italy}
\affiliation{\it Department of Physics, University of Cincinnati, Cincinnati, Ohio 45221, USA}
\affiliation{\it LIP Coimbra, P-3004-516 Coimbra,  Portugal}
\affiliation{\it Niels Bohr Institutet, DK 2100 Copenhagen, Denmark}
\affiliation{\it Institut f\"ur Physik, Universit\"at Dortmund, D-44221 Dortmund, Germany}
\affiliation{\it Joint Institute for Nuclear Research Dubna, 141980 Dubna, Moscow region, Russia}
\affiliation{\it DESY, D-22603 Hamburg, Germany}
\affiliation{\it Max-Planck-Institut f\"ur Kernphysik, D-69117 Heidelberg, Germany}
\affiliation{\it Physikalisches Institut, Universit\"at Heidelberg, D-69120 Heidelberg, Germany}
\affiliation{\it Department of Physics, University of Houston, Houston, TX 77204, USA}
\affiliation{\it Institute for Nuclear Research, Ukrainian Academy of Science, 03680 Kiev, Ukraine}
\affiliation{\it J.~Stefan Institute, 1001 Ljubljana, Slovenia}
\affiliation{\it University of Ljubljana, 1001 Ljubljana, Slovenia}
\affiliation{\it University of California, Los Angeles, CA 90024, USA}
\affiliation{\it Lehrstuhl f\"ur Informatik V, Universit\"at Mannheim, D-68131 Mannheim, Germany}
\affiliation{\it University of Maribor, 2000 Maribor, Slovenia}
\affiliation{\it Institute of Theoretical and Experimental Physics, 117259 Moscow, Russia}
\affiliation{\it Max-Planck-Institut f\"ur Physik, Werner-Heisenberg-Institut, D-80805 M\"unchen, Germany}
\affiliation{\it Dept. of Physics, University of Oslo, N-0316 Oslo, Norway}
\affiliation{\it Fachbereich Physik, Universit\"at Rostock, D-18051 Rostock, Germany}
\affiliation{\it Fachbereich Physik, Universit\"at Siegen, D-57068 Siegen, Germany}
\affiliation{\it Institute for Nuclear Research, INRNE-BAS, Sofia, Bulgaria}
\affiliation{\it Universiteit Utrecht/NIKHEF, 3584 CB Utrecht, The Netherlands}
\affiliation{\it DESY, D-15738 Zeuthen, Germany}
\affiliation{\it Physik-Institut, Universit\"at Z\"urich, CH-8057 Z\"urich, Switzerland}
\affiliation{\it visitor from Dipartimento di Energetica dell' Universit\`{a} di Firenze and INFN Sezione di Bologna, Italy}
\affiliation{\it visitor from P.N.~Lebedev Physical Institute, 117924 Moscow B-333, Russia}
\affiliation{\it visitor from Moscow Physical Engineering Institute, 115409 Moscow, Russia}
\affiliation{\it visitor from Moscow State University, 119899 Moscow, Russia}
\affiliation{\it visitor from Institute for High Energy Physics, Protvino, Russia}
\affiliation{\it visitor from High Energy Physics Institute, 380086 Tbilisi, Georgia}
\author{I.~Abt} 
\affiliation{\it Max-Planck-Institut f\"ur Physik, Werner-Heisenberg-Institut, D-80805 M\"unchen, Germany}
\author{M.~Adams} 
\affiliation{\it Institut f\"ur Physik, Universit\"at Dortmund, D-44221 Dortmund, Germany}
\author{M.~Agari}
\affiliation{\it Max-Planck-Institut f\"ur Kernphysik, D-69117 Heidelberg, Germany}
\author{H.~Albrecht} 
\affiliation{\it DESY, D-22603 Hamburg, Germany}
\author{A.~Aleksandrov} 
\affiliation{\it DESY, D-15738 Zeuthen, Germany}
\author{V.~Amaral} 
\affiliation{\it LIP Coimbra, P-3004-516 Coimbra,  Portugal}
\author{A.~Amorim} 
\affiliation{\it LIP Coimbra, P-3004-516 Coimbra,  Portugal}
\author{S.~J.~Aplin} 
\affiliation{\it DESY, D-22603 Hamburg, Germany}
\author{V.~Aushev} 
\affiliation{\it Institute for Nuclear Research, Ukrainian Academy of Science, 03680 Kiev, Ukraine}
\author{Y.~Bagaturia} 
\affiliation{\it DESY, D-22603 Hamburg, Germany}
\affiliation{\it visitor from High Energy Physics Institute, 380086 Tbilisi, Georgia}
% ...

\collaboration{HERA-B Collaboration}

\noaffiliation

\date{\today}

\begin{abstract}
We have searched for \thp(1540) and \ximm(1862) pentaquark 
candidates in proton-induced reactions on C, Ti and W targets 
at mid-rapidity and $\sqrt{s} = 41.6$~GeV. In $2\cdot10^8$ 
inelastic events we find no evidence for  narrow ($\sigma\approx5$\,\MEV )
signals in the \thp$\rightarrow$p\ks\ and \ximm$\rightarrow$\xim\pim\ channels;
our 95\% CL upper limits (UL)   for the inclusive 
production cross section times branching fraction \BRS $|_{y\approx0}$  are 3.7 and 2.5~$\mu$b/N. 
The UL of the yield ratio of \thp\,/\,\lam(1520)$<$2.7\,\%
is significantly lower than model predictions. Our UL of \BRD \ximm\,/\,\xisz $<$4\% 
is at variance with the results that have provided first evidence 
for the \ximm\ signal.   
\end{abstract}

\pacs{14.20.Jn, 13.85.Rm, 12.39-x, 12.40-y}

\maketitle



Recent experimental evidence suggests not only that pentaquarks (PQs), 
i.e. baryons with at least five constituent quarks, exist but that their production in
high energy collisions is common.
After the possible discovery of the \thp\ PQ ($uudd\bar{s}$) at 1540~\MEV\ in 
the $\gamma$n$\rightarrow$K$^-$K$^+$n process on carbon \cite{leps}, more than 10 experiments using 
incident beams of photons, electrons, kaons, protons or (anti)-neutrinos have 
observed resonances within $\pm$20~\MEV\ of this mass in either the n\kp\ 
\cite{r2-4} or 
the p\ks\ 
\cite{r5-9,zeus,svd} decay channels; 
the measured widths have all been consistent with the experimental resolutions ranging from 
20~\MEV\ to 2~\MEV\,\cite{zeus}.
The \thp\ interpretation is based on a prediction \cite{dia97} of the chiral soliton model (CSM)
according to which the \thp\ is expected to have a mass of 1530~\MEV , a width of less than 15~\MEV , 
and to decay into the KN channel. 
In both the CSM and the correlated quark model \cite{jaf03}, the \thp\ is a member of an antidecuplet 
with two further exotic isospin 3/2 states of $S=-2$, the \ximm\ ($ddss\bar{u}$) and the 
$\Xi^+_{3/2}$ ($uuss\bar{d}$). 
In pp collisions at $\sqrt{s}\approx18$~GeV, narrow candidate resonances for both 
the \ximm\ and its neutral isospin partner have been found in 
the \xim\pim\ and \xim\pip\ final states at the mass of 1862~\MEV\ \cite{na49}. 
Theoretically, PQs are not restricted to the strange sector, and 
experimental evidence for an anti-charmed  PQ, \thc\ ($uudd\bar{c}$), with a mass of 
3.1~\GEV\ has recently been reported \cite{h1}. 
In this context also earlier already `forgotten' c$\bar{\rm{c}}$ PQ candidates 
\cite{karn92} have been recalled \cite{mihu04}.

On the other hand, criticism addressed to some of the  reported PQ signals includes the problem 
of kinematic reflections \cite{dzi03}, of spurious states \cite{zav03}, and of low statistics 
\cite{fis04}. 
Other puzzles include the surprisingly narrow width of the \thp\ \cite{cah03}, the large and 
systematic \cite{zhao04} spread of measured \thp\ masses, and the non-observation of the \thc\ in 
an equivalent experiment \cite{nothc}.
Hence, for establishing the existence and character of the new resonances, 
high statistics mass spectra are needed as well as measurements of spin, parity, width and cross 
sections.
In addition, considering the results of high statistics  studies which have found neither the \thp\ 
signal in $\psi(2S)$ and $J/\psi$ hadronic decays \cite{bes} nor
the \ximm\ signal in $\Sigma^-$-induced reactions on nuclear targets \cite{wa89}, the need
for a thorough understanding of the PQ production mechanism has been emphasized \cite{kar04}.
Benchmarks for PQ production exist based on statistical hadronization models; they typically
predict particle ratios such  as \thp /\lam(1520) in heavy ion \cite{ran03,let03,bec03} and 
pp \cite{bec03,liu04,wer04ble04} collisions. 
Taking advantage of a large data sample with good mass resolution (see Table~\ref{tab:t1})
HERA-B can contribute significantly to many of these topics.
The simultaneous study of 
\thp\,$\rightarrow$\,p\ks\,$\rightarrow$p\pim\pip\ and 
\ximm\,$\rightarrow$\,\xim\pim\,$\rightarrow$\lam\pim\pim\
decays in 
proton-nucleus collisions at $\sqrt{s}$=41.6~GeV allows a test of these  theoretical
predictions and a comparison with earlier experimental results including
the possible first confirmation of the \ximm\  signal.
\begin{table}
\caption{\label{tab:t1} Statistics  and experimental resolutions $\sigma$ of 
the relevant signals (charge-conjugate modes indicated by c.c.). 
} 
\begin{ruledtabular}
\begin{tabular}{lccc}
Signal         & C target & all targets & $\sigma$/(\MEV ) \\
\hline
  \ks                    & 2.2M          &   4.9M      & 4.9  \\
\lam\ [c.c.]              & 440k\,[210k]    &  1.1M\,[520k] & 1.6  \\
\lam(1520) [c.c.]         & 1.3k\,[760]    &  3.5k\,[2.1k] & 2.3     \\
\xim\ [c.c.]              & 4.7k~[3.4k]   & 12k~[8.2k]  & 2.6  \\
\xisz\ [c.c.]             & 610~[380]     & 1.4k~[940] & 2.9  \\
\end{tabular}
\end{ruledtabular}
\end{table}
 
HERA-B is a fixed target experiment at the 920 GeV proton storage ring of DESY. 
It is a forward magnetic spectrometer with a large acceptance centered at mid-rapidity 
(y$_{cm}\approx0$), 
featuring a high-resolution vertexing and tracking system and excellent particle identification 
\cite{abt03}.
The present study is based on a sample of $2\cdot10^8$ 
minimum bias events which were recorded at $\sqrt{s}=41.6$~GeV  
using carbon (C), titanium (Ti) and tungsten (W) wire targets during the 2002/03 run period.
For this analysis the information from the silicon vertex detector, the main tracking system,
the ring-imaging Cherenkov counter (RICH), and the electromagnetic calorimeter (ECAL) was used.

With standard techniques described in \cite{abt03}, signals from \ks~$\rightarrow$~\pip\pim ,
\lam~$\rightarrow$~p\pim\ and \alam~$\rightarrow$~$\bar{\rm p}$\pip\ decays are identified above
a small background without particle identification (PID) requirements.
\begin{figure}
\begin{center}
\includegraphics[width=8.5cm]{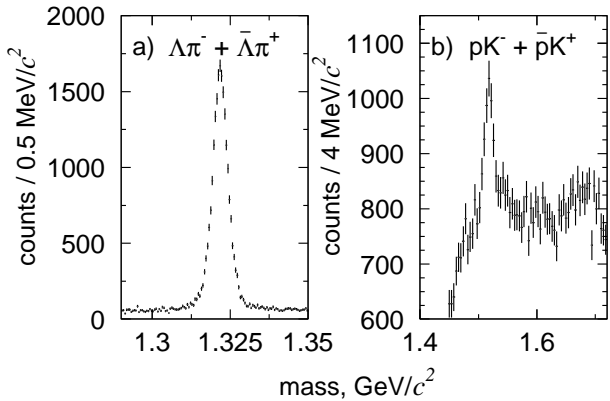}
\caption{\label{fig:d1} Signals obtained with the C target from decays of
         a) \xim$\rightarrow$\lam\pim\,and \xip$\rightarrow$\alam\pip , 
	 and 
         b) $\Lambda(1520)\rightarrow$\,pK$^-$ and \alam(1520)$\rightarrow\bar{\rm p}$K$^+$. 
}		  
\end{center}
\end{figure}
Similar clean signals from \xim~$\rightarrow$~\lam\pim\ and c.c. decays (Fig.~\ref{fig:d1}a) are 
obtained by requesting  the \lam\pim\ vertex to be at least 2.5~cm downstream of the target and 
the event to exhibit a cascade topology: a further downstream \lam\ vertex and  the \xim\ 
pointing back to the target wire (impact parameter $b<1$~mm). 
Table~\ref{tab:t1} summarizes the statistics of these signals, together 
with their measured mass resolutions $\sigma$.  These resolutions  are about 20\% 
larger than those of the Monte Carlo (MC) simulation, while all  mass values agree within $<$1\,\MEV\ with 
the nominal masses. 
For all particle selections, invariant masses are required to be  within $\pm3\sigma$ of the 
respective nominal mass.

\begin{figure}[h]
\centering
\begin{center}
\includegraphics[width=8.5cm]{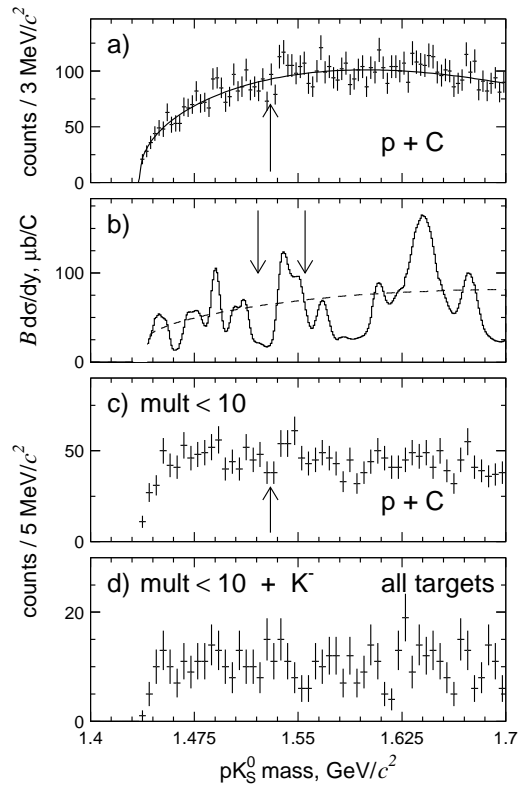}
\caption{\label{fig:d2} 
The pK$^0_S$ invariant mass distributions: 
a) data from the p+C collisions and the background estimate (continuous line); 
b) deduced UL(95\%)  for the p+C inclusive cross section 
   at mid-rapidity; the dashed line shows our 95\% CL sensitivity; 
c,d) same as a) but requiring c) a track multiplicity of $<$10, and d) in addition a K$^-$ particle 
in the event.
The arrows mark the masses of 1521, 1530 and 1555~\MEV .}
\end{center}
\end{figure}

For the search for \thp\,$\rightarrow$\,p\ks\ decays, events with at least one reconstructed primary
vertex were selected. The proton 
PID was provided by the RICH. 
The cut in  proton likelihood of $>0.95$ implies a misidentification probability of less than 1\% 
in the selected momentum range from 22 to 55~GeV/{\it c} 
\cite{RICH}.
The \lam\ and \alam\ contaminations \cite{zav03} were removed \cite{abt03} in the \ks \ sample.
The invariant mass spectrum of the p\ks\ pairs  is shown in 
Fig.~\ref{fig:d2}a) for the p+C data.
The solid line represents the background determined from event mixing after normalization to the data. 
The spectrum exhibits a smooth shape in the mass region from 1.45 to 1.7~\GEV . 
Using the prescription of ref.~\cite{F+C},
we have calculated from these data upper limits at 95\% confidence level, UL(95\%), for the inclusive 
production cross section of a narrow resonance at mid-rapidity, \BRS $|_{y\approx0}$, 
(Fig.~\ref{fig:d2}b); the y$_{cm}$-interval is $\pm0.3$.
The data have been fitted with a Gaussian plus a background of fixed shape.
The mean of the Gaussian was varied in steps of 1~\MEV\ but fixed in the fit;
its width was fixed to the MC prediction multiplied by 1.2 and increased 
from 2.6 to 6.1 \MEV\ over the considered range. At the \thp\ mass, the width was 3.9~\MEV .
The reconstruction efficiencies have been determined by MC simulations assuming a flat
rapidity distribution and a $p_t^2$ distribution proportional to $\exp(-B\cdot p_t^2)$ with 
$B=2.1$\,(GeV/\sol )$^{-2}$
\cite{abt03}.
Assuming an atomic mass dependence of A$^{0.7}$ for the production cross section, the UL(95\%)
of \BRS\ varies from 3 to 22~$\mu$b/nucleon (N) for a \thp\ mass  between 1521 and 1555~\MEV .
A systematic error of 14\% was taken into account.
For the \thp\ mass of 1530~\MEV (about the average of the mass values observed in the
p\ks\ final state \cite{zhao04}), our limit is \BRS$\,<\,3.7$~$\mu$b/N.  
The ULs from all target data are within $\pm30$\% of these values.

Further search strategies were tried including 
i) a cut on the track multiplicity of the event (Fig.~\ref{fig:d2}c) which would otherwise peak at $\approx13$,
ii) the request of a tagging particle such as a \lam , $\Sigma$ or \km\ in the event, or 
iii) both conditions (Fig.~\ref{fig:d2}d). 
None yielded a statistically significant 
structure in \thp\ mass region.
Also, the effect of lowering the cut on the RICH proton likelihood and the corresponding
increase of the proton momentum acceptance has been systematically studied without yielding a
\thp\ signal. 
On the other hand, as shown in Fig.~\ref{fig:d1}b, when the same proton PID requirement used to
produce Fig.~\ref{fig:d2} is applied to p\km\ candidates, a strong \lam(1520) signal results,
further demonstrating the capabilities of the RICH. The cut in the \km\ likelihood of $>0.95$
implies a selection of kaon momenta  from 12 to 55~GeV/{\it c}. 
With the same cut on the \ks\ momenta,
and assuming a branching ratio of 
\BR(\thp$\rightarrow$p\ks)\,=\,0.25, the UL(95\%) of the particle ratio 
\thp$(1530)$/\lam(1520) at y$_{cm}\approx0$ is 2.7\%. 

\begin{figure}[h]
\begin{center}
\includegraphics[width=8.5cm]{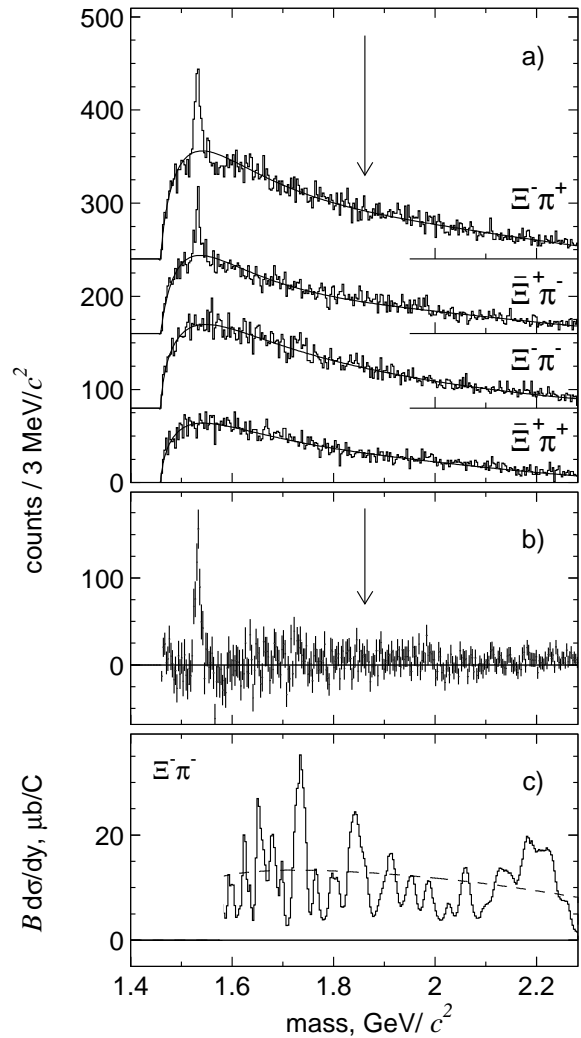}
\caption{\label{fig:d4} 
The $\Xi \pi$ invariant mass distributions:
a) data from the p+C collisions in indicated neutral and doubly-charged channels and the background 
estimates (continuous lines);
b) sum of all four $\Xi \pi$ spectra with the background subtracted, and 
c) deduced UL(95\%)  for the p+C inclusive cross section at mid-rapidity. The dashed line shows our 
95\% CL sensitivity.
The arrows mark the mass of 1862~\MEV .}
\end{center}
\end{figure}

Both doubly-charged and neutral $\Xi_{3/2}$ PQ candidates as well as their anti-particles have been 
searched for in the $\Xi\,\pi$ channels.
The pion candidates were required to originate from the primary vertex.
The background was further reduced by weak cuts on the PIDs from the ECAL and RICH which
eliminated all the tracks with a positive electron, proton, or kaon PID.  
The histograms of Fig.~\ref{fig:d4}a) show the resulting 
$\Xi\,\pi$ invariant mass spectra obtained from the C target data. 
The smooth lines are the background estimates from event-mixing normalized to the data.
In the neutral channels  the \xisz\ resonance shows up as a prominent signal
of $\approx 10^3$ events (see Table~\ref{tab:t1}). 
The observed width ($\approx$9.5~\MEV ) of the \xisz\ agrees  with MC
simulations which imply an experimental resolution  of 2.9~\MEV .
None of these mass spectra shows evidence for the narrow, less than 18~\MEV\ (FWHM) wide 
PQ candidates at 1862~\MEV\ reported by the NA49 collaboration \cite{na49} nor for any
other narrow state at masses between 1.6 and 2.3~\GEV .
Fig.~\ref{fig:d4}b) shows the sum of the four spectra of Fig.~\ref{fig:d4}a) after 
background subtraction and can be compared directly to  Fig.~3 of ref.~\cite{na49}.
The corresponding ULs(95\%) of the production cross sections \BRS $|_{y\approx0}$ per
carbon nucleus at mid-rapidity (Fig.~\ref{fig:d4}c) have been obtained in the same way as 
described above 
for the p\ks\ channel; here the y$_{cm}$-interval is $\pm0.7$, the experimental resolution increases 
from 2.9 to 10.6~\MEV\ in the considered mass range, and  is 6.6~\MEV\ at 1862~\MEV . 
At this \xim\pim\,mass, the UL(95\%) of \BRS\ is 2.5~$\mu$b/N; the corresponding 
limits in the \xim\pip, \xip\pip, and \xip\pim channels are 2.3, 0.85, and 3.1~$\mu$b/N.
With an A$^{0.7}$ dependence, the ULs from all targets are 2.7, 3.2, 0.94, 
and 3.1~$\mu$b/N, respectively. 

Table~\ref{tab:t2} lists our  ULs(95\%) of various 
relative yields for the \thp\ and \ximm . Reference states for the \thp\ are
the \lam\ and the \lam(1520), and for the \ximm , the \xim\ and the
\xisz . The \thp\ and \ximm\ widths are assumed to be equal to our experimental 
mass resolution and their momentum distributions are assumed to be
the same as those of the reference states.
Table~\ref{tab:t2} lists also predictions of various statistical hadronization 
models for the respective ratios.
We note that these ratios show no significant variation between  $17<\sqrt{s}<40$~GeV,
nor is there a significant difference between predictions for pp and AA collisions.
We find  our UL for \thp /\lam(1520)$<$2.7\% to be more
than one order of magnitude lower than the model predictions. 
Also, the UL of \thp /\lam $<0.92$\% is lower than all predictions 
including the model which
uses the Gribov-Regge approach for describing the \thp\ production and its $\sqrt{s}$ dependence 
in pp collisions \cite{wer04ble04}. 
Our UL of the \ximm /\xim\ yield ratio is compatible with the model predictions.
No theoretical value is yet available for the \ximm /\xisz\ ratio, but our UL of 
$<$4\%/\BR\
should be compared with the value from the NA49 experiment which, however, is not
explicitly quoted in the original paper \cite{na49} which 
reports only the number of 38 \ximm\ events. 
According to ref. \cite{fis04}, the number of \xisz\ events is about 150 
leading to a yield ratio \cite{kad04} in contradiction to our UL
unless the relative efficiencies for \xisz\
and \ximm\ of NA49 (unpublished)  differ markedly from those of HERA-B.
\begin{table}
\caption{\label{tab:t2}
Our 95\% CL upper limits on the relative yields of \thp(1530) and \ximm(1862) PQs at y$_{cm}\approx0$ 
and predictions for pp and AA collisions. 
For a \thp\ mass of 1540~\MEV , the quoted values have to be multiplied by $\approx$4. 
}
\begin{ruledtabular}
\begin{tabular}{lcccccc}
Reaction & $\sqrt{s_{NN}}$ & $ \Theta^+ \over \Lambda $   & $ \Theta^+ \over \Lambda(1520)$  & $\Xi^{--} \over\Xi^-$ & $\Xi^{--} \over\Xi(1530)^0$ & Ref.\\
         &    [GeV]   &     [\%]   &     [\% ]    &     [\% ]        &     [\% ]  \\        
\hline
pA, y$\approx0$  & 42 & $<0.92$       &   $<2.7$       &   $<3$/\BR    & $<$4/\BR       & \\
	   \hline
pp, y=0        & 18 & 2.3          &              &                  & & \cite{wer04ble04}  \\
pp             & 20/40 & 6.3/5.0          &              &     2.5/3.6          & &  \cite{liu04} \\
pp             & 17 & 4.7          &    57       &                   & &  \cite{bec03} \\
AA             & 20 & 3-10          &  50-200     &    0.4-1         & &  \cite{let03} \\
               & 40 & 3-7          &  44-140     &     0.4-1         & &  \cite{let03} \\
\end{tabular}
\end{ruledtabular}
\end{table}

In conclusion, having found no evidence for narrow \thp\ and \ximm\  signals,
we have set UL(95\%) for the central production cross sections of  
resonances in the p\ks\ and \xim\pim\ final states with widths less than our
experimental resolution of $\approx$\,5\,\MEV .
For the \thp(1530) and the \ximm(1862) the respective ULs
of \BRS $|_{y\approx0}$ are 3.7 and 2.5~$\mu$b/N.  
For the \thp\ candidate observed in pA collisions at $\sqrt{s}=11.5$\,~GeV, the total cross 
section
for x$_F\ge0$ was estimated to be 30 to 120~$\mu$b/N \cite{svd}. 
A decrease of the central \thp\ production with increasing $\sqrt{s}$ could be understood
if the \thp\ is produced by  disintegration of forward/backward peaked remnants~\cite{wer04ble04}. 
On the other hand, our UL(95\%) for \thp /\lam(1520)$<2.7$\% is significantly lower 
than statistical hadronization predictions which yield a ratio of $\geq$0.5  in agreement 
with experiments in which  the \thp\ candidate and \lam(1520) showed similar yields
\footnote{Upon completion of this paper, we learned that the SPHINX collaboration 
has searched for the \thp\ in exclusive proton-induced reactions on carbon at $\sqrt{s}=11.5$~GeV 
studying  four different final states of the \thp $\bar{\rm K}^0$ system. No evidence
for a narrow PQ peak is found in any of the studied channels \cite{sphi}.}.
Our UL(95\%) of \BRD \ximm /\xim $<$3\% is not low enough to contradict the 
theoretical predictions. It is, however, inconsistent with the previously-published \cite{na49}  
observation of the \ximm(1862) at mid-rapidity which is based on a data sample of 
lower statistics (1.6k~v.~12k\,\xim ) and comparable mass resolution (7.6~v.~6.6\,\MEV ). 

\begin{acknowledgments}
The collaborating institutions wish to thank DESY for its support and kind hospitality.
This work is supported by
NSRC (Denmark);
BMBF, DFG, and MPRA (Germany);
INFN (Italy);
FOM (The Netherlands);
RC (Norway);
POCTI (Portugal);
MIST (No.\,SS1722.2003, Russia);
MESS (Slovenia);
CICYT (Spain);
SNF (Switzerland);
NAS and MES (Ukraine);
DOE and NSF (U.S.A.);
\end{acknowledgments}

\end{document}